\documentclass[10pt]{article}
\def\paradot#1{\vspace{1.3ex plus 0.7ex minus 0.5ex}\noindent{\bf\boldmath{#1.}}}
\def\paradot#1{\vspace{1.3ex plus 0.7ex minus 0.5ex}\noindent{\bf\boldmath{#1.}}}
\usepackage{natbib}
\usepackage{latexsym,amsmath,amssymb,amsthm,amsfonts,graphicx}
\include{def}
\def\,{\mskip 3mu} \def\>{\mskip 4mu plus 2mu minus 4mu} \def\;{\mskip 5mu plus 5mu} \def\!{\mskip-3mu}
\def\dispmuskip{\thinmuskip= 3mu plus 0mu minus 2mu \medmuskip=  4mu plus 2mu minus 2mu \thickmuskip=5mu plus 5mu minus 2mu}
\def\textmuskip{\thinmuskip= 0mu                    \medmuskip=  1mu plus 1mu minus 1mu \thickmuskip=2mu plus 3mu minus 1mu}
\textmuskip
\def\be{\dispmuskip\begin{equation}}    \def\ee{\end{equation}\textmuskip}
\def\beqn{\dispmuskip\begin{displaymath}}\def\eeqn{\end{displaymath}\textmuskip}
\def\bea{\dispmuskip\begin{eqnarray}}    \def\eea{\end{eqnarray}\textmuskip}
\def\bqan{\dispmuskip\begin{eqnarray*}}  \def\eqan{\end{eqnarray*}\textmuskip}

\def\v{\boldsymbol}
\def\SetR{I\!\!R}

\def\I{1\!\!1} 				% identity matrix
\def\v{\boldsymbol}

\def\SetR{I\!\!R}

\def\a{\alpha}
\def\D{\Delta}

\def\g{\gamma}
\def\s{\sigma}
\def\t{\theta}
\def\b{\beta}
\def\l{\lambda}

\def\X{{\cal X}}
\def\Y{{\cal Y}}

\def\PPS{\text{\rm PPS}}
\def\KL{\text{\rm KL}}

\def\OLS{\text{\rm OLS}}

\def\diag{\text{\rm diag}}

\def\plasso{\text{\rm plasso}}
\def\wplasso{\text{\rm wplasso}}

\begin{document}

\title{The Predictive Lasso}
\author{{ Minh-Ngoc Tran\footnote{Corresponding author: Department of Statistics and Applied Probability, National University of Singapore, Singapore 117546.
E-mail: ngoctm@nus.edu.sg.}, David J. Nott, Chenlei Leng}\\
\normalsize Department of Statistics and Applied Probability\\
\normalsize National University of Singapore}
\date{\empty}
\maketitle
  %=======
\begin{abstract}
We propose a shrinkage procedure for simultaneous variable selection and estimation
in generalized linear models (GLMs) with an explicit predictive motivation.
The procedure estimates the coefficients by minimizing the Kullback-Leibler divergence
of a set of predictive distributions to the corresponding predictive distributions
for the full model, subject to an $l_1$ constraint on the coefficient vector.  
This results in selection of a parsimonious model with similar predictive performance
to the full model.  Thanks to its similar form to the original lasso problem for GLMs,
our procedure can benefit from available $l_1$-regularization path algorithms.
Simulation studies and real-data examples confirm the efficiency of our method 
in terms of predictive performance on future observations. 

\paradot{Keywords}
Generalized linear models,
Kullback-Leibler divergence,
Lasso,
Optimal prediction,
Variable selection.
\end{abstract}  
%============================================================================%  
\section{Introduction}\label{secIntro}
%============================================================================%

A primary goal in statistics is to develop algorithms that predict future data well from past 
observations.  In regression problems where a large number of predictors are involved
predictive accuracy in statistical modeling may depend to a large extent on model selection strategies.  
For generalized linear models (GLMs), for example,
a large number of potential predictors are often given in order to reduce modeling bias, and 
one then would like to select a smaller subset achieving some kind of optimality properties.
Popular methods such as the lasso and its variants can achieve model selection consistency, i.e., if the true model 
was included in the model set under consideration,
these methods would be able to identify (asymptotically) the true model.
However, whether or not the true model exists is a controversial issue.
For a real dataset, it is believed either that no true model exists
or that the true model has an infinite number of parameters (Burnham and Anderson, 2002).
In this paper we deal with the problem of estimation and variable selection
for GLMs with the goal of prediction in mind.

From the Bayesian perspective it is sometimes argued that the full model should be used 
to achieve the best prediction accuracy (Aitchison, 1975; Geisser, 1993).  
However, with many predictors prior specification and elicitation may be difficult, 
and the full model does not have {\em interpretability} - 
a property that is often desirable for many statistical procedures -
because it does not tell us which and how predictors affect the response.
Another drawback of using the full model is that if there is a cost
associated with data collection then it would be inadvisable to use all of the predictors.
This motivates the idea of choosing a submodel
whose predictive distribution is close to that of the full model.
This idea has been somewhat recognized in the literature. 
Brown et al. (2002) look at Bayesian model averaging incorporating variable selection for prediction.
Tran (2009) and Vehtari and Lampinen (2004) propose model selection methods based
on Kullback-Leibler divergence from the predictive distribution of the full model to the predictive
distributions of the submodels.   
These works are motivated by the idea of trading off between prediction accuracy and parsimony.
However, these methods are challenging to implement
because searching over the whole model space is computationally infeasible.
Like the idea of the lasso (Tibshirani, 1996),
we overcome this problem by using $l_1$ constraints on the coefficients.
By doing this, we can enjoy the computational advantages of the algorithms for convex optimization with $l_1$ constraints. 
Unlike the lasso, however, our approach has an explicit predictive motivation
which aims at selecting a useful model with high prediction accuracy.
A related approach is considered by Nott and Leng (2010) based on Kullback-Leibler projections, 
motivated by earlier work of Dupuis and Robert (2003) although these approaches
are not based directly on posterior predictive distributions.  

For a collection of $N$ predictive distributions obtained from the full model, we write
$\KL_i(M_\text{full}\|M_\b)$, $i=1,...,N$ for the Kullback-Leibler divergences from 
the predictive distributions of the model based on coefficient vector $\v\b$
to those of the full model.  Our
approach in its general form is to solve for $\v\b$ the following optimization problem 
\beqn
\min_\b\sum_{i=1}^N \KL_i(M_\text{full}\|M_\b) + \lambda\|\v\b\|_{l_1}
\eeqn
with $\l$ a shrinkage parameter as in the original lasso.
The main contribution of the present paper is to motivate and develop such a procedure for 
variable selection and estimation in GLMs that
(i) automatically simultaneously estimates the coefficients and selects significant predictors;
(ii) achieves good prediction accuracy;
(iii) is broadly applicable;
(iv) is computationally efficient. 
This procedure will be called {\it the predictive lasso} or plasso for short.

The plasso for GLMs will be presented in Section \ref{secplasso}.
Section \ref{secPrior} presents useful prior specifications which can
facilitate computation. In particular, 
we discuss in more detail the plasso for linear models
and extend our previous discussion to a weighted version of the basic approach.
Simulation and real-data examples are presented in Section \ref{secExp} to demonstrate the use of the plasso
and to compare it with the adaptive lasso (Zou, 2006) in terms of predictive performance.
%Section \ref{secDiscussion} discusses various extensions of the plasso.
Section \ref{secConclusion} contains concluding remarks.

%============================================================================%  
\section{The predictive lasso}\label{secplasso}
%============================================================================%

We consider the problem of estimation and variable selection for GLMs
with potential covariates $\v x=(x_0\equiv1,x_1,...,x_p)'\in\X$ and the response $y\in\Y$.
With a suitable link function $g$, $g(E(y|\v x))$ is assumed to be a linear combination of $\v x$
\be\label{appro}
g(E(y|\v x))=\b_0+\b_1x_1+...+\b_px_p=\v x'\v\b.
\ee
We assume that the covariates $x_i$ are in their final forms, 
no further transformations are needed
(i.e., for various reasons and in order to keep things simple,
we restrict ourselves to the linear approximation \eqref{appro}).
The sampling distribution of an observation $\D_i=(\v x,\ y)$ then
is assumed to have the following form
\beqn
p(\D |\v\b,\phi)=p(\v x)p(y|\v x,\v\b,\phi)\propto p(\v x)\exp\left(\dfrac{1}{a(\phi)}\big[y_i\t(\v x'\v\b)-b(\t(\v x'\v\b))\big]\right),
\eeqn
where $\v\b\in\SetR^{p+1},\ \phi>0$ are the coefficient vector and scale parameter, respectively, and 
$\t,\ a$ and $\ b$ are known functions.
If covariates $\v x$ are not random, the density $p(\v x)$ in the above expression can be omitted.
We are concerned with the problem of simultaneous coefficient estimation and variable selection
with the goal of prediction in mind. 
Like the lasso, we would like to develop a method 
for simultaneous variable selection and parameter estimation.
However, unlike the lasso our approach has a more explicit predictive motivation,
which aims at producing a useful model with high prediction accuracy.

Given the past dataset $D$ and certain priors for parameters $(\v\b,\ \phi)$ of the full model,
the predictive distribution $p(\Delta|D)$ for a future observation $\Delta=(\v x,y)$ is given by
\be\label{pre.dist.}
p(\D|D)=p(\v x|D)p(y|\v x,D)=p(\v x|D)\int p(y|\v x,\v\b,\phi)p(\v\b,\phi|D)d\v\b d\phi.
\ee
We can assume that $p(\v x|D)\equiv p(\v x)$, i.e., future design points are independent of past data.
We propose to estimate the coefficient vector $\v\b$ by solving the following optimization problem:
\be\label{plasso}
\min_{\b}\int\log\dfrac{p(\Delta|D)}{p(\Delta|\v\b,\phi)}p(\Delta|D)d\Delta
%\int_\X\int_\Y \log\dfrac{p(y|\v x,D)}{p(y|\v x,\v\b,\phi)}p(y|\v x,D)dyd\v x
\;\;\;\;\text{s.t.}\;\;\;\;\sum_{j=1}^pw_j|\beta_j|\leq\tau
\ee
where the tuning parameter $\tau\geq0$ and weights $w_j\geq0$ are chosen later.
(As will become clear shortly, $\phi$ plays no role in this optimization problem, we can assume at the moment that $\phi$ is known).
Note that the objective function is the Kullback-Leibler divergence 
from $p(\D|\v\b,\phi)$ to the predictive distribution $p(\D|D)$.
We refer to this procedure of estimating $\v\b$ through the optimization of \eqref{plasso} as the predictive lasso (plasso).  

Let $\{\D_t=(\v x_t,y_t),\ t=1,...,T\}$ be Markov chain Monte Carlo (MCMC) samples from the predictive distribution $p(\D|D)$.
The integral in \eqref{plasso} then can be approximated by the average $(1/T)\sum_{t=1}^T\log[{p(\D_t|D)}/{p(\D_t|\v\b,\phi)}]$,
and \eqref{plasso} becomes
\be\label{plasso1}
\min-\frac1T\sum_{t=1}^T\log{p(\D_t|\v\b,\phi)}\;\;\;\;\text{s.t.}\;\;\;\;\sum_{j=1}^pw_j|\beta_j|\leq\tau,
\ee
or more specifically
\be\label{plasso2}
\min\frac1T\sum_{t=1}^T\big[b(\t(\v x_t'\v\b))-y_t\t(\v x_t'\v\b)\big]\;\;\;\;\text{s.t.}\;\;\;\;\sum_{j=1}^pw_j|\beta_j|\leq\tau.
\ee
This optimization problem is also equivalent to
\be\label{plasso3}
\min\frac1T\sum_{t=1}^T\big[b(\t(\v x_t'\v\b))-y_t\t(\v x_t'\v\b)\big]+\l\sum_{j=1}^pw_j|\beta_j|
\ee
where $\l$ is a tuning parameter.
Such an optimization problem is easier to deal with if 
the objective function is convex.
The convexity of the objective function turns out to depend on the link function, and holds 
for most popular GLMs.

Often, the integral in $\v x$ is approximated by a sum over $N$ points $\v x_1^f,...,\v x_N^f$.
These points might not coincide with the observed design points, they ``come from the future" 
(hence the superscript ``f" stands for ``future").
For each $\v x_i^f$, let $\bar y_i^f$ be the mean of MCMC samples $\{y_{it},t=1,...,T_0\}$ drawn from $p(y_i^f|\v x_i^f,D)$ - 
the predictive distribution of the future response $y_i^f$ at design point $\v x_i^f$ given past data $D$.
Then, it is easy to see that \eqref{plasso3} becomes
\be\label{plasso4}
\min\frac1N\sum_{i=1}^N[b(\t(\v\b'\v x_i^f))-\bar y_i^f\t(\v\b'\v x_i^f)]+\l\sum_{j=1}^pw_j|\beta_j|.
\ee
Note that, under the squared error loss, $\bar y_i^f$ is an estimate of the best prediction 
(w.r.t. the predictive distribution $p(y_i^f|\v x_i^f,D)$) for the response at $\v x_i^f$.
As will be seen in Section \ref{secPrior}, for linear regression with a convenient specification of priors there is 
no need to conduct MCMC because the predictions $\bar y_i^f=E(y_i^f|\v x_i^f,D)$ have a closed form.

We have approximated the integral over $\v x$ by a sum over $N$ ``future" points $\v x_i^f,\ i=1,...,N$.
Typically, these points are specified depending on the context and/or 
on the distribution $p(\v x)$ over $\X$.
%For example, if we want to construct a model for prediction with predictor points within only a sub-range $\X'\subset\X$,
%then the points $\v x_i^f$ may be correspondingly drawn from $\X'$.
As a default implementation of our procedure, however, 
we propose to identify the future points $\v x_i^f$ with the observed training points $\v x_i,\ i=1,...,n$.
The reason behind this is that if the sample size $n$ is large enough and the observed training points $\v x_i$ were
randomly selected from $p(\v x)$, then
by the law of large numbers the integral over $\v x$ can be well approximated by the sum over $\v x_i$.
In what follows therefore, if not otherwise specified, we consider the plasso for GLMs in the following form
\be\label{plasso5}
\min\frac1n\sum_{i=1}^n[b(\t(\v x_i'\v\b))-\bar y_i^f\t(\v x_i'\v\b)]+\l\sum_{j=1}^pw_j|\beta_j|.
\ee
Note that the original (adaptive) lasso for GLMs is
\be\label{olasso}
\min\frac1n\sum_{i=1}^n[b(\t(\v x_i'\v\b))-y_i\t(\v x_i'\v\b)]+\l\sum_{j=1}^pw_j|\beta_j|.
\ee
The plasso in this form differs from the original lasso only in the way
it replaces the observed responses $y_i$ by the predictions $\bar y_i^f=E(y_i^f|\v x_i,D)$.
Available routines to solve \eqref{olasso} then can be used for \eqref{plasso5}.

We have not yet considered the issue of choice of the tuning parameters in
the plasso.  As the primary goal of the plasso is to predict the future, 
cross-validation is a very natural choice for estimating $\l$. 
As in the adaptive lasso, the weights $w_j$ can be assigned as $1/|\tilde \b_j|$
with $\tilde\b_j$ the MLE of $\b_j$.
In a Bayesian context it is also natural to consider $\tilde{\beta}_j$ as the posterior mode.

%============================================================================%  
\section{Some useful prior specifications}\label{secPrior}
%============================================================================%
Given the available routines to solve the optimization problem of form \eqref{plasso5},
all what we need to implement the plasso is to calculate the quantities 
$\bar y_i^f=E(y_i^f|\v x_i,D)$. 
To do so, in general, we first need to specify a useful prior for parameters, determine posterior distributions
and then estimate $\bar y_i^f=E(y_i^f|\v x_i,D)$ by MCMC or some other method.
However, in some cases there is no need to conduct MCMC.   
We first present in this section a prior specification for linear models 
in which the predictions $\bar y_i^f$ have closed form.
For genalized linear models, we present here two prior specifications.
The first is adapted from \cite{Chen:2003} which is interpretable in terms of observables rather than parameters.
The second one proposed recently by \cite{Gelman:2008} is useful for routine applied use.

%============================================================================%  
\subsection{Prior specification for linear models}\label{linear.plasso}
%============================================================================%
Consider the linear model
\beqn
\v y=X\v\beta+\v\epsilon
\eeqn
where $\v y$ is the $n$-vector of responses, $X$ is an $n\times (p+1)$ design matrix
%where rows are observations and columns are variables 
and $\v\epsilon$ is an $n$-vector of iid normal errors with mean zero and variance $\sigma^2$.  
The $(p+1)$-vector $\v\beta$ consists of unknown mean parameters and 
we consider the situation where $\sigma^2$ is also unknown.  
Consider the conjugate prior specification (O'Hagan and Forster, 2004, Chapter 11) 
$p(\v\beta,\sigma^2)=p(\sigma^2)p(\v\beta|\sigma^2)$
in which $p(\sigma^2)$ is inverse gamma 
\beqn
p(\s^2) = \frac{(a/2)^{(d/2)}}{\Gamma(d/2)}(\s^2)^{-d/2-1}\exp(-\frac{a}{2\s^2})
\eeqn
and $p(\v\beta|\sigma^2)$ is multivariate
normal, $N(\v m,\sigma^2 V)$.  With these priors the predictive distribution of a new observation $\D=(\v x,y)$
is $p(\D|D)=p(\v x|D)p(y|\v x,D)$ with $p(y|\v x,D)=t_{d+n}\left(\v x'\tilde{\v\beta},s^2(1+\v x' \hat{V} \v x)\right)$
where
\bqan
\tilde{\v\beta}&=&(X'X+V^{-1})^{-1}(V^{-1}\v m+X'\v y),\\
\hat V&=&(V^{-1}+X'X)^{-1},\\
%s^2&=&\frac{a+(\v y-X\hat{\v\beta})'(\v y-X\hat{\v\beta})+(\v m-\hat{\v\beta})'(V+(X'X)^{-1})^{-1}(\v m-\hat{\v\beta})}{n+d-2},\\
s^2&=&\frac{a+\v m'V^{-1}\v m+\v y'\v y-(V^{-1}\v m+X'\v y)'(V^{-1}+X'X)^{-1}(V^{-1}\v m+X'\v y)}{n+d-2},\\
\hat{\v\beta}&=&(X'X)^{-1}X'\v y.
\eqan
We write $w(\v x)=1+{\v x}' \hat{V} \v x$.  

Now consider the predictive lasso \eqref{plasso} where as usual the integral over $\v x$ 
is approximated by a sum over $N$ ``future" points $\v x_i^f$. 
Then equivalently, we need to minimize (the scale $\phi$ is now re-denoted by $\sigma^2$)
\be\label{lplasso}
\sum_{i=1}^N\int \left[-\log p(y_i^f|\v x_i^f,\v\b,\sigma^2)\right]p(y_i^f|\v x_i^f,D)dy_i^f\;\;\;\text{s.t.}\;\;\sum_{j=1}^p w_j|\beta_j|\leq \tau.
\ee
Noting that 
\beqn
\log p(y_i^f|\v x_i^f,\v\beta,\sigma^2)=-\frac{1}{2}\log 2\pi\sigma^2-\frac{1}{2\sigma^2}(y_i^f-(\v x_i^f)'\v\beta)^2,
\eeqn
minimizing \eqref{lplasso} is equivalent to minimizing
\be\label{lplasso1}
\frac{N}{2}\log \sigma^2+\frac{1}{2\sigma^2}\sum_{i=1}^N E\left((y_i^f-(\v x_i^f)'\v\beta)^2|\v x_i^f,D\right)\;\;\;\text{s.t.}\;\;\sum_{j=1}^p w_j|\beta_j|\leq \tau.
\ee
With the closed form of the predictive distribution as a $t$-distribution 
we have
\beqn
E\left((y_i^f-(\v x_i^f)'\v\beta)^2|\v x_i^f,D\right)=s^2 w(\v x_i^f)+\left((\v x_i^f)'\tilde{\v\beta}-(\v x_i^f)'\v\beta\right)^2.
\eeqn
Substituting this into \eqref{lplasso1} we must minimize
\be\label{lplasso2}
\frac{N}{2}\log \sigma^2+\frac{1}{2\sigma^2}\sum_{i=1}^N s^2 w(\v x_i^f)
+\frac{1}{2\sigma^2} \sum_{i=1}^N \left((\v x_i^f)'\tilde{\v\beta}-(\v x_i^f)'\v\beta\right)^2
\ee
subject to the constraint.  Minimizing this as a function of $\v\beta$ amounts as before to an ordinary
lasso problem where the responses are replaced with the fitted values from the full model
at the future design points $\v x_i^f$, $i=1,...,N$.  With a non-informative prior and with the $\v x_i^f$
as the observed design points $\v x_i$ this is the ordinary lasso, since in this case $\tilde{\v\beta}=\hat{\v\beta}$
and for the least squares estimator 
\beqn
\sum_{i=1}^n (y_i-\v x_i'{\v\beta})^2=\sum_{i=1}^n (y_i-\v x_i'\hat{\v\beta})^2+
\sum_{i=1}^n (\v x_i'\hat{\v\beta}-\v x_i'\v\beta)^2
\eeqn
where the first term on the right hand side does not depend on $\v\beta$.  

If \eqref{lplasso2} has been minimized with respect to $\v\beta$ subject to the constraint to
obtain an estimate $\hat{\v\beta}_\plasso$ (this in general depends on the constraint $\tau$ but we
suppress this in the notation) then substituting in $\hat{\v\beta}_\plasso$ and minimizing with
respect to $\sigma^2$ gives
\bea
\hat{\sigma}^2_\plasso & = & \frac{\sum_{i=1}^N \mbox{Var}(y_i^f|\v x_i^f,D)+\sum_{i=1}^N \left((\v x_i^f)'\tilde{\v\beta}-(\v x_i^f)'\hat{\v\beta}_\plasso\right)^2}
{N}\notag \\
& = & \frac{\sum_{i=1}^N s^2 w(\v x_i^f) +\sum_{i=1}^N \left((\v x_i^f)'\tilde{\v\beta}-(\v x_i^f)'\hat{\v\beta}_\plasso\right)^2}
{N}.
\eea

%------------------------------------------%
\paradot{The weighted version of plasso}
%------------------------------------------%
One extension we can consider that gives different results to the ordinary lasso in the noninformative
case with the $\v x_i^f$ the observed design points $\v x_i$ is the following.  Suppose that instead of considering
predictive distributions in our predictive lasso objective function 
where the variance does not depend on $\v x$ we predict $y_i^f$ with
\beqn
p(y_i^f|\v\beta,\sigma^2 w(\v x_i^f))=N(\v (x_i^f)'\v\beta,\sigma^2 w(\v x_i^f))
\eeqn
That is, we allow our normal form predictive distributions to have variances which vary in proportion
to the true predictive variances in the full model $\mbox{Var}(y_i^f|\v x_i^f,D)$.
The standard deviation in the full model $\sqrt{\mbox{Var}(y_i^f|\v x_i^f,D)}$ 
is often considered a more realistic estimate of the standard error,
because it incorporates model uncertainty.
We now consider minimization of
\beqn
\sum_{i=1}^N \int \left[-\log p(y_i^f|\v\beta,\sigma^2 w(\v x_i^f))\right] p(y_i^f|\v x_i^f,D)dy_i^f
\eeqn
subject to the constraint and following a similar argument to our previous one we must minimize
\beqn
\sum_{i=1}^N \frac{1}{w(\v x_i^f)} \left((\v x_i^f)'\tilde{\v\b}-(\v x_i^f)'\v\beta\right)^2
\eeqn
subject to the constraint in order to estimate $\v\beta$.  This is similar to before, but now
with weights of $1/w(\v x_i^f)$ for the different design points.  
We will refer to this producdure as the weighted plasso (wplasso).
After $\v\beta$ has
been estimated as $\hat{\v\beta}_\wplasso$ say, the minimization with respect to $\sigma^2$
gives
\begin{eqnarray*}
\hat{\sigma}^2_\wplasso & = & \frac{\sum_{i=1}^N \frac{1}{w(x_i)}\mbox{Var}(y_i^f|\v x_i^f,D)+\sum_{i=1}^N \frac{1}{w(x_i^f)}\left((\v x_i^f)'\tilde{\v\beta}-(\v x_i^f)'\tilde{\v\beta}_\wplasso\right)^2}
{N} \\
& = & \frac{\sum_{i=1}^N s^2  +\sum_{i=1}^n \frac{1}{w(x_i^f)}\left((\v x_i^f)'\tilde{\v\beta}-(\v x_i^f)'\tilde{\v\beta}_\wplasso\right)^2}
{N}.
\end{eqnarray*}

%------------------------------------------%
\paradot{Elicitation of hyperparameters}
%------------------------------------------%
We now discuss on the choice of the hyperparameters $\v m$ and $V$.
There are many different ways proposed for choosing the matrix $V$ in the literature.
For example, \cite{Zellner:1986} proposed the so-called {\em g-prior}
in which $V$ is set equal to $c(X'X)^{-1}$ with some $c>0$
($c=n$ is a common choice).
\cite{Raftery:1997} proposed an alternative where $V$ is a block-diagonal matrix.
For noncategorical covariates, $V$ is a diagonal matrix $\diag(s_y^2,\kappa^2s_1^{-2},...,\kappa^2s_p^{-2})$
where $s_y^2$ is the sample variance of $\v y$, and 
$s_i^2$ are the variances of the columns of $X$.
For a categorical covariate, the corresponding diagonal element will be a matrix induced from the corresponding dummy variables.
\cite{Raftery:1997} proposed a value of $2.85$ for $\kappa$ together with $a=0.72$ and $d=2.58$.
For the parameter $\v m$, they proposed the default value of $\v m=(\hat\b_0^\OLS,0,...,0)'$
where $\hat\b_0^\OLS$ is the OLS estimate of $\b_0$.
An alternative is $\v m=\v 0$. 
These two choices of $\v m$ often lead to very similar inferences.
We will use the setup of \cite{Raftery:1997} in our following numerical examples.

%============================================================================%  
\subsection{Prior specifications for generalized linear models}\label{GLM.plasso}
%============================================================================%
There is an extensive literature on prior specifications for GLMs.
We will briefly present here two of them: the first one is due to \cite{Chen:2003}
and the second is proposed recently by \cite{Gelman:2008}.  

%-----------------------------------%
\paradot{The Chen and Ibrahim prior}
%-----------------------------------%
Recall that the sampling distribution of observables $\v y=(y_1,...,y_n)$ in the GLM case is
\beqn
p(\v y|X,\v\b,\phi)\propto\exp\left(\sum_1^n\dfrac{1}{a(\phi)}\big[y_i\t(\v x_i'\v\b)-b(\t(\v x_i'\v\b))\big]\right)=\exp\left(\dfrac{1}{a(\phi)}\big[\v y'\v\t-\I'\v b(\v\t)\big]\right)
\eeqn
where $\v\t=\v\t(\v\b)=(\t_1,...,\t_n)',\ \t_i=\t(\v x_i'\v\b),\ \v b(\v\t)=(b(\t_1),...,b(\t_n))'$ and $\I$ is an $n-$vector of 1s.
For ease of exposition, we assume that $\phi$ is known 
(and therefore suppressed in the notation), as, for example, in logistic and Poisson regression.
\cite{Chen:2003} proposed the following prior for $\v\b$
\be\label{prior}
p(\v\b)\propto\exp\left(\g_0\dfrac{1}{a(\phi)}\big[\v\a_{0}'\v\t-\I'\v b(\v\t)\big]\right)
\ee
where $\g_0\geq0$ and $\v\a_0\in\SetR^n$ are hyperparameters determined later on.
Denote this distribution by $\v\b|\phi\sim D(\g_0,\v\a_0)$. 
They proved that the prior \eqref{prior} is proper and that 
this prior is conjugate with the posterior $\v\b|X,\v y\sim D(1+\g_0,(\g_0\v\a_0+\v y)/(1+\g_0))$.

As shown by \cite{Chen:2003}, $E(\v y)=\v\a_0$,
it is natural to choose $\v\a_0$ as a prior guess for $E(\v y)$.
Therefore, in practice, $\v\a_0$ should be obtained from experts in the field
although default empirical Bayes alternatives such as choosing $\v\a_0$ as the fitted values based on
the MLE or other methods are also possible.
The parameter $\g_0$ weighs the importance of the prior guess.
In general, $\g_0$ should be taken such that $\g_0=\g_0(n)\to0$ as $n\to\infty$, 
i.e., the prior has less influence when more data is available. 
An advantage of this prior specification is that it is interpretable in terms of observables
rather than parameters which are sometimes not easy to elicit. 

%-----------------------------------%
\paradot{The Gelman et al. prior}
%-----------------------------------%
\cite{Gelman:2008} proposed a weakly informative prior distribution for GLMs,
constructed by first standardizing the covariates to have mean zero and standard deviation 0.5,
and then putting independent $t-$distributions on the coefficients.
As a default choice, they recommended a central Cauchy distribution with scale 10 for the intercept
and central Cauchy distributions with scale 2.5 for other coefficients.
As argued by \cite{Gelman:2008}, 
this prior specification has many advantages;
besides, it works in an automatic fashion with no hyperparameter elicitation needed.

Recall that all what we need to implement the plasso is to calculate the quantities $\bar y_i^f=E(y_i^f|\v x_i,D)$. 
After the prior has been specified, $\bar y_i^f$ can be estimated 
by MCMC or some other method.
It is well-known that 
\beqn
E(\v y|X,\v\b)=\dot{\v b}(\v\t)=(\dot b(\t_1),...,\dot b(\t_n))',
\eeqn
so that
\be\label{post3}
\bar{\v y}^f=E(\v y^f|X,\v y)=E_{\b|X,y}\left[E(\v y^f|X,\v\b)\right]=E_{\b|X,y}[\dot{\v b}(\v\t(\v\b))]
\ee
which can be easily estimated by MCMC samples from the posterior distribution $\v\b|X,\v y$.

A procedure for fitting GLMs with the Gelman et al. prior has been implemented in R.  
In the following numerical examples for logistic regression where no expert advice is available,
we use the default prior of Gelman et al.
For high-dimensional cases where using MCMC may be time consuming, we suggest using
the plug-in predictive density to estimate the predictions $\bar y_i^f$. 
Our experiences show that this is very fast compared to MCMC. 

%============================================================================%  
\section{Experiments}\label{secExp}
%============================================================================%
In this section, we study the plasso through simulations and real-data examples. 
We use the convenient prior specifications as in Section \ref{secPrior}.  
The tuning parameter $\l$ is selected by 5-fold cross-validation. 
The examples are carried out using R with the help of the R packages glmnet and arm.

A popular measure of predictive ability is the {\em partial predictive
score} (PPS) (Good, 1952; Geisser, 1980; Hoeting et al., 1999). Suppose that the
data is split into two parts, the training set $D^T$ and the prediction set $D^P$ . 
The partial predictive score of the distributions induced by model parameters $(\v\b^*,\phi^*)$ is defined as
\beqn
\PPS=-\frac{1}{|D^P|}\sum_{\Delta=(x,y)\in D^P}\log p(y|\v x,\v\b^*,\phi^*).
\eeqn
It is understood that smaller PPS means better predictive performance.

\paradot{Example 1: A simulation study for linear regression} 
Consider the following linear model
\be\label{linear}
y = 2 + \v x'\v\b+\sigma\epsilon
\ee 
where $\v\b=(3,\ 2,\ 0,\ 0,\ 0.5,\ 0.5,\ 0,\ 0)'$
(so that there are some main and also small effects), $\epsilon$ is iid $N(0,1)$,
and $\sigma>0$ is the noise level.
We want to compare the predictive performance of the plasso and the wplasso to that of the adaptive lasso (alasso).
In our first simulation study, design points $\v x_j$ are simulated from
a multivariate normal distribution $N_8(\v 0,\Sigma)$ with $\sigma_{ij}=0.5^{|i-j|}$.
We first generate from model \eqref{linear} a dataset which serves as the training set $D^T$.
Another dataset $D^P$ then is generated, which is used to test the predictive performance.
Table \ref{linearsimulation} presents the PPS (after ignoring the constants independent of models) 
and numbers of zero-estimated coefficients averaged over 500 replications with
various factors $n^T$ (size of training set), $n^P$ (size of prediction set) and $\sigma$.
In each case, the first row gives PPS and the second the numbers of zero-estimated coefficients.
The numbers in parentheses are standard deviations.
The results suggest that the plasso and wplasso have better predictive ability than the alasso.
As one may expect for predictively motivated methods, 
models selected by the plasso and wplasso are less sparse than selected by the alasso. 

In our second simulation study, design points $\v x_j$ are simulated from a multivariate $t$-distribution 
with degrees of freedom being 1.5.
By doing so, we intend to simulate situations in which some predictors have high leverage points,
i.e. their distributions have long tails.
The simulation result is presented in Table \ref{longtail}.  
As one may expect, the wplasso works better and more stable than the orthers
because the variance is modeled to vary in proportion to the true predictive variance.   
    
In our last simulation study, we try a high-dimensional example.
We consider the linear model \eqref{linear} with $p=100$ and most of the coefficients are zero 
except $\beta_j=5$, $j=10,20,...,100$.
The result reported in Table \ref{largep} suggests that in general the plasso and wplasso
compare favourably with the alasso in this example.  

 \begin{table}
  \begin{center}
    \begin{tabular}{c|c|c|c|c}
%&&\multicolumn{2}{c|}{small-$p$ case}&\multicolumn{2}{c}{large-$p$ case}\\
$n_T=n_P$&$\sigma$&alasso&plasso&wplasso\\
\hline 
50	&	1	&	0.7775(0.1684)	&	0.6066(0.1395)&	0.6100(0.1290)  \\
	&		&	4.0200(0.9281)  &	2.3440(1.2201)&2.5240(1.1472)	\\
\cline{2-5}
	&	2	&	1.4592(0.1830)	&	1.3006(0.1535)&	1.2949(0.1422)	\\
	&		&	5.3000(0.9674)  &	2.8800(1.3643)& 3.1760(1.3344)	\\
\cline{2-5}
	&	3	&	1.8555(0.1613)	&	1.6968(0.1390)&	1.6922(0.1317)	\\
	&		&	5.7560(0.7600)	&	3.3900(1.4457)&	3.6540(1.4051)	\\
\hline
100	&	1	&	0.6871(0.1143)	&	0.5536(0.0842)&	0.5544(0.0828)	\\
	&		&	3.8200(0.6069)  &	2.2440(1.1537)& 2.3340(1.1424)	\\
\cline{2-5}
	&	2	&	1.3713(0.1056)	&	1.2458(0.0857)&	1.2463(0.0845)	\\
	&		&	4.9920(0.9432)	&	2.5260(1.3239)&	2.6180(1.2975)	\\
\cline{2-5}
	&	3	&	1.7719(0.1110)	&	1.6488(0.0840)&	1.6478(0.0830)	\\
	&		&	5.5660(0.7687)	&	2.8120(1.4481)&	2.9320(1.4337)	\\
\hline
200	&	1	&	0.6252(0.0721)	&	0.5226(0.0525)&	0.5234(0.0524)	\\
	&		&	3.8360(0.4260)	&	2.1020(1.1446)&	2.1740(1.1447)	\\
\cline{2-5}
	&	2	&	1.3199(0.0739)	&	1.2197(0.0576)&	1.2197(0.0575)	\\
	&		&	4.5200(0.8503)	&	2.3100(1.1561)&	2.3820(1.1518)	\\
\cline{2-5}	
	&	3	&	1.7236(0.0688)	&	1.6244(0.0547)&	1.6243(0.0545)	\\
	&		&	5.4600(0.7833)	&	2.4760(1.2508)&	2.5280(1.2411)
   \end{tabular}
  \end{center}
  \caption{PPS and numbers of zero-estimated coefficients averaged over 500 replications with normal predictors. The numbers in parentheses are standard deviations.}\label{linearsimulation}
\end{table}

 \begin{table}
  \begin{center}
    \begin{tabular}{c|c|c|c|c}
%&&\multicolumn{2}{c|}{small-$p$ case}&\multicolumn{2}{c}{large-$p$ case}\\
$n_T=n_P$&$\sigma$&alasso&plasso&wplasso\\
\hline 
50	&	1	&	5.1060(35.720)	&	2.3525(13.262)&	0.7268(0.2619)  \\
	&		&	2.9860(1.1010)  &	1.7120(1.2280)& 2.0340(1.1999)	\\
\cline{2-5}
	&	2	&	4.6046(21.094)	&	2.2795(4.8080)&	1.3623(0.2467)	\\
	&		&	3.5340(1.1898)  &	1.9140(1.1702)& 2.2560(1.2319)	\\
\cline{2-5}
	&	3	&	4.4344(22.570)	&	3.5535(23.491)&	1.7629(0.2079)	\\
	&		&	4.0260(1.4260)	&	2.3320(1.3555)&	2.6360(1.3383)	\\
\hline
100	&	1	&	3.3895(39.616)	&	0.9106(2.2240)&	0.6270(0.1969)	\\
	&		&	3.0860(0.9963)  &	1.6100(1.2136)& 1.8360(1.2329)	\\
\cline{2-5}
	&	2	&	3.5369(20.497)	&	1.4987(0.8000)&	1.2759(0.1244)	\\
	&		&	3.2380(1.0195)	&	1.8480(1.1966)&	2.0860(1.2139)	\\
\cline{2-5}
	&	3	&	2.6614(3.5284)	&	3.4847(29.991)&	1.6794(0.1497)	\\
	&		&	3.4680(1.1899)	&	1.9880(1.2405)&	2.1920(1.2158)	\\
\hline
200	&	1	&	1.6475(10.968)	&	0.7445(1.1904)&	0.5858(0.2033)	\\
	&		&	3.1900(0.9569)	&	1.3600(1.2703)&	1.6740(1.2533)	\\
\cline{2-5}
	&	2	&	1.9182(4.1911)	&	1.5461(3.0734)&	1.2510(0.1717)	\\
	&		&	3.0960(1.0024)	&	1.6020(1.2010)&	1.8660(1.1979)	\\
\cline{2-5}	
	&	3	&	2.3556(4.5199)	&	1.8363(1.2353)&	1.6384(0.0662)	\\
	&		&	3.3240(1.0262)	&	1.8640(1.2134)&	 2.1400(1.1896)
   \end{tabular}
  \end{center}
  \caption{PPS and numbers of zero-estimated coefficients averaged over 500 replications with long-tailed $t$-distribution predictors. The numbers in parentheses are standard deviations.}\label{longtail}
\end{table}
      
\begin{table}
  \begin{center}
    \begin{tabular}{c|c|c|c|c}
%&&\multicolumn{2}{c|}{small-$p$ case}&\multicolumn{2}{c}{large-$p$ case}\\
$n_T=n_P$&$\sigma$&alasso&plasso&wplasso\\
\hline 
50	&	1	&	6.6472(51.647)	&	3.4224(6.9586)&	2.5732(0.1725)  \\
	&		&	76.440(5.0352)  &	61.940(7.2292)& 68.320(8.3188)	\\
\cline{2-5}
	&	2	&	8.9814(51.873)	&	7.0791(20.430)&	2.9250(0.4229)	\\
	&		&	83.940(9.4159)  &	73.085(10.190)& 80.805(9.8439)	\\
\hline
100	&	1	&	1.3247(1.1160)	&	1.0808(1.3710)&	1.0307(0.0861)	\\
	&		&	77.105(19.757)  &	61.465(17.118)& 74.315(10.851)	\\
\cline{2-5}
	&	2	&	2.1953(2.7703)	&	2.2110(2.5382)&	1.6085(0.1005)	\\
	&		&	74.525(19.389)	&	61.160(19.573)&	74.240(11.700)	\\
\hline
200	&	1	&	0.9484(0.1250)	&	0.6047(0.0651)&	0.6717(0.0504)	\\
	&		&	89.155(1.0802)	&	66.420(9.7709)&	75.235(6.2975)	\\
\cline{2-5}
	&	2	&	1.6169(0.1059)	&	1.3039(0.0662)&	1.3182(0.0505)	\\
	&		&	89.110(1.1064)	&	66.650(9.8158)&	75.300(6.1325)	\\
   \end{tabular}
  \end{center}
  \caption{PPS and numbers of zero-estimated coefficients averaged over 500 replications for the large-$p$ case. The numbers in parentheses are standard deviations.}\label{largep}
\end{table}

\paradot{Example 2: Linear regression - predicting percent body fat}
Percentage of body fat is one important measure of health,
which can be accurately estimated by underwater weighing techniques \citep{Bailey:1994}.
These techniques often require special equipment and are sometimes not convenient,
thus fitting percent body fat to simple body measurements is a convenient way
to predict body fat. 
\cite{Johnson:1996} introduced a dataset in which
percent body fat and 13 simple body measurements (such as weight, height and abdomen circumference) 
are recorded for 252 men.
After omitting observations 39 (because a weight value of 363.15 pounds is unusually large),
42 (because a height value of 29.5 inches is unreasonable), and
182 (because the response value is 0),
we obtain a dataset of size 249.

We are concerned with the problem of constructing
a model that predicts the response from the covariates. 
Following Hoeting et al. (1999), we use a linear regression model.
The primary goal is prediction accuracy for future observations; 
besides this, parsimony is another important objective, since a simple model
is preferred for the sake of scientific insight into the $x - y$ relationship.

Using the full dataset, the alasso, plasso and wplasso estimates of $\v\b$
are given in Table \ref{table1}.
These methods simultaneously do parameter estimation and variable selection,
because some of the estimated coefficients are exact zero.
Recall that the goals at which the methods aim are somewhat different:
plasso and wplasso have a more explicit predictive motivation;
besides, the wplasso in some cases is somewhat more realistic in the sense that it allows
the variances to vary in proportion to the predictive variance of the full model.

 \begin{table}
  \begin{center}
    \begin{tabular}{|c|rrr|rrr|rrr|rrr|}
\hline
&\multicolumn{3}{|c|}{full data}&\multicolumn{3}{c|}{case I}&\multicolumn{3}{c|}{case II}&\multicolumn{3}{c|}{case III}\\
&al  &pl	 &   wpl&	al  &pl &wpl& al&pl &wpl&al &pl &wpl	\\	
\hline
$C$&-18.00	&    6.79&     -0.18&   -14.78	&2.88	&-0.28	&-15.69	&-2.95	&-4.59	&-23.31	&-0.61	&-3.87	\\
$X_1$&0	&    0.06&	0.04&	0.02	&0.09	&0.08	&0	&0	&0	&0	&0	&0	\\
$X_2$&0	&       0&	   0&	0	&0	&0	&0	&0	&0	&0	&0	&0	\\
$X_3$&-0.20	&   -0.29&     -0.27&	-0.26	&-0.40	&-0.39	&0	&-0.17	&-0.14	&0	&-0.24	&-0.22	\\
$X_4$&0	&   -0.30&     -0.11&	0	&-0.24	&-0.17	&0	&0	&0	&0	&-0.34	&-0.25	\\
$X_5$&0	&   -0.09&      0   &	0	&0	&0	&0	&0	&0	&0	&0	&0	\\
$X_6$&0.55	&    0.78&	0.68&	0.55	&0.70	&0.68	&0.38	&0.66	&0.66	&0.45	&0.69	&0.69	\\
$X_7$&0	&   -0.09&      0   &	0	&-0.09  &0	&0	&0	&0	&0	&0	&0	\\
$X_8$&0	&    0.09&	0   &	0	&0.16   &0.08	&0	&0	&0	&0	&0	&0	\\
$X_9$&0	&       0&	   0&	0	&0.09	&0	&0	&0	&0	&0	&0	&0	\\
$X_{10}$&0	&    0.09&	   0&	0	&0.22	&0.17	&0	&-0.39	&-0.43	&0	&-0.04	&0	\\
$X_{11}$&0	&    0.13&	0.04&	0	&0 	&0	&0	&0.10	&0.10	&0	&0.20	&0.20	\\
$X_{12}$&0	&    0.19&	0   &	0	&0	&0	&0	&0 	&0	&0	&0.19	&0.07	\\
$X_{13}$&0	&   -1.62&     -1.31&	0	&-1.34	&-1.20	&0	&-1.16	&-1.15	&0	&-1.44	&-1.35	\\
\hline
PPS&	&	 &	    &	1.946	&1.933	&1.933	&2.112	&1.913	&1.902	&2.075	&1.965	&1.951	\\		 	
\hline
   \end{tabular}
  \end{center}
  \caption{Predicting percent body fat. The abbreviations ``al", ``pl" and ``wpl" stand for
alasso, plasso and wplasso, respectively.}\label{table1}
\end{table}

\iffalse
 \begin{table}
  \begin{center}
    \begin{tabular}{|rrr|rrr|rrr|rrr|}
\hline
\multicolumn{3}{|c|}{full data}&\multicolumn{3}{c|}{case I}&\multicolumn{3}{c|}{case II}&\multicolumn{3}{c|}{case III}\\
al  &pl	 &   wpl&	al  &pl &wpl& al&pl &wpl&al &pl &wpl	\\	
\hline
-18.00	&    6.79&     -0.18&   -14.78	&2.88	&-0.28	&-15.69	&-2.95	&-4.59	&-23.31	&-0.61	&-3.87	\\
0	&    0.06&	0.04&	0.02	&0.09	&0.08	&0	&0	&0	&0	&0	&0	\\
0	&       0&	   0&	0	&0	&0	&0	&0	&0	&0	&0	&0	\\
-0.20	&   -0.29&     -0.27&	-0.26	&-0.40	&-0.39	&0	&-0.17	&-0.14	&0	&-0.24	&-0.22	\\
0	&   -0.30&     -0.11&	0	&-0.24	&-0.17	&0	&0	&0	&0	&-0.34	&-0.25	\\
0	&   -0.09&      0   &	0	&0	&0	&0	&0	&0	&0	&0	&0	\\
0.55	&    0.78&	0.68&	0.55	&0.70	&0.68	&0.38	&0.66	&0.66	&0.45	&0.69	&0.69	\\
0	&   -0.09&      0   &	0	&-0.09  &0	&0	&0	&0	&0	&0	&0	\\
0	&    0.09&	0   &	0	&0.16   &0.08	&0	&0	&0	&0	&0	&0	\\
0	&       0&	   0&	0	&0.09	&0	&0	&0	&0	&0	&0	&0	\\
0	&    0.09&	   0&	0	&0.22	&0.17	&0	&-0.39	&-0.43	&0	&-0.04	&0	\\
0	&    0.13&	0.04&	0	&0 	&0	&0	&0.10	&0.10	&0	&0.20	&0.20	\\
0	&    0.19&	0   &	0	&0	&0	&0	&0 	&0	&0	&0.19	&0.07	\\
0	&   -1.62&     -1.31&	0	&-1.34	&-1.20	&0	&-1.16	&-1.15	&0	&-1.44	&-1.35	\\
\hline
	&	 &	    &	1.946	&1.933	&1.933	&2.112	&1.913	&1.902	&2.075	&1.965	&1.951	\\		 	
\hline
   \end{tabular}
  \end{center}
  \caption{Predicting percent body fat. Abbreviations ``al", ``pl" and ``wpl" stand for
alasso, plasso and wplasso, respectively.}\label{table1}
\end{table}
\fi

We now examine the predictive performance of these three procedures. To this end, we split
the dataset into two parts: the first 125 observations are used as the
training set $D$, the remaining observations are used as the prediction set $D^P$.
The alasso, plasso and wplasso estimates and their PPS are given in Table \ref{table1} (case I).
As the second examination,  the first 125 observations are used as the
prediction set $D^P$, the remaining observations are used as the training set $D$.
For the third examination, we randomly split the full dataset into two (roughly) equal parts
which serve as the training and prediction sets. 
The estimates and PPS are summarized in Table \ref{table1}. 
As one may expect for predictively motivated methods, 
the variables selected by plasso and wplasso in general contain those selected by alasso,
i.e., the models selected by plasso and wplasso are bigger than the one selected by alasso. 
In all cases, the plasso and wplasso show a better predictive performance over the alasso.  
It seems that modelling the variances to 
vary in proportion to the predictive variance of the full model
is appropriate in this example,
because the wplasso has a similar or better predictive performance compared with the plasso. 

\paradot{Example 3: Logistic regression - the Chapman data}
This dataset consists of 6 features of 200 men as well as binary responses indicating the presence of heart disease.
This dataset is described in detail in \cite{Christensen:1997}.
We use logistic regression to explore the relationship between the presence of heart disease and the features.    

Table \ref{table4} presents the result of analysis.
The quantities $\bar y_i^f$ are estimated from MCMC samples as in \eqref{post3}
with 10000 draws after discarding 1000 burn-in samples. 
To examine the predictive performance, we use the first half of observations as the training set, the rest as the prediction set.
In the second case, we use the last 100 observations as the training set and the first 100 for testing of predictive performance;
and for the last case 100 observations are randomly selected for the training set.
In all cases, as shown in Table \ref{table4}, plasso shows a better predictive performance over alasso.      

 \begin{table}
  \begin{center}
    \begin{tabular}{|c|rr|rr|rr|rr|}
\hline
&\multicolumn{2}{c|}{full data}&\multicolumn{2}{c|}{case I}&\multicolumn{2}{c|}{case II}&\multicolumn{2}{c|}{case III}\\
&alasso&plasso&alasso&plasso&alasso&plasso&alasso&plasso	\\	
\hline 
intercept&	-7.86&	-5.10&	-1.73&	-8.63&	-2.92&	-0.87&	-4.29&	-3.42\\
$X_1$ &     	0.05 &	0.04 &	0    &	0.04 &	0.02 &	0.05 &	0.02 &   0.04\\
$X_2$ &   	0    &  0    &	0    &  0.01 &  0    &  0    &  0    &   0\\
$X_3$ &		0    &	0    &	0    &	0    &  0    &  -0.01&	0    &	 0\\
$X_4$ &       	0    &	0    &	0    &	0    &	0    &	0    &  0    &   0\\
$X_5$ &   	0    &  -0.06&  0    & -0.03 &  0    & -0.08 &  0    &   -0.08\\
$X_6$ &   	0.01 &  0.02 &  0    &  0.02 &	0    &	0.01 &  0.01 &   0.02\\
\hline
  PPS &	   	     &	     &0.3533 &0.3233 &0.4125 &0.3762 &0.3452 &   0.3302\\
\hline
   \end{tabular}
  \end{center}
  \caption{Example 3: the Chapman data.}\label{table4}
\end{table}

\paradot{Example 4: Logistic regression - the spambase data}
We consider in this example another application of the predictive lasso in the logistic regression framework
that is more challenging with many predictors and instances.
We consider the spam email data set created by Mark Hopkins, Erik Reeber, George Forman and Jaap Suermondt
at the Hewlett-Packard Labs.
The data set consists of 4061 messages,
each has been already classified as email or spam
together with 57 attributes (predictors) which are relative frequencies of commonly occurring words.
The goal is to design a spam filter that
could filter out spam before clogging the users' mailboxes.
Our goal as usual is to construct a parsimonious model with a good prediction accuracy.

With a large number of predictors and observations, using MCMC may be time consuming
so that we use the plug in method discussed earlier.
We randomly split the data set into two parts (training set and prediction set).
Both alasso and plasso select a subset of 34 predictors out of 57 with 
the PPS are 0.2604 and 0.2538 respectively.
In our second try, the alasso selects 28 predictors and gives a PPS of 0.2710
while the plasso selects 30 predictors and gives a PPS of 0.2562. 
Repeat this many times, we observe that the plasso often select several predictors 
more than the alasso and that the plasso {\em almost always} has a smaller PPS.  

%============================================================================%  
\section{Conclusion}\label{secConclusion}
%============================================================================%
The popular lasso as a procedure for simultaneous variable selection and estimation has many attractive properties,
and under certain conditions is able to identify the true model if it is assumed to exist.
Our suggested plasso has a more explicit predictive motivation 
which aims at selecting a useful model for prediction;
besides, it enjoys the attractive properties of lasso.
A notable feature of plasso is that we put no restriction on the predictive distribution $p(\Delta|D)$.
Although we have considered $p(\Delta|D)$ as arising from a full model
including all potential covariates, it can in fact arise from any model where a 
GLM approximation with variable selection is desired.  The approximation
can also be an appropriately local one in the covariate space through a 
judicious choice of the design points in the plasso criterion, which
need not correspond to the observed design points.  
In this paper we have motivated and developed the idea of plasso only for GLMs.
It is clear that this idea can be extended to other models rather than GLMs, and this
is a topic for future research.

\bibliographystyle{apalike}

\end{document}